\documentclass[aps,pra,twocolumn,showpacs,groupedaddress,superscriptaddress]{revtex4-1}
\usepackage{bm,graphicx,amsmath}
\usepackage[utf8]{inputenc}
\usepackage[T1]{fontenc}
\usepackage{lmodern}
\usepackage{amssymb}
\usepackage{amsmath}
\usepackage{upgreek}
\usepackage{epsfig}
\usepackage{natbib}
\usepackage[dvipsnames]{xcolor}
\usepackage{makecell}
\usepackage{bm}
\usepackage[pdftex,colorlinks,linkcolor=blue,citecolor=blue,urlcolor=blue]{hyperref}
\usepackage{array}
\usepackage{multirow} 
\usepackage{booktabs}
\usepackage[normalem]{ulem}

\newcommand{\ket}[1]{\ensuremath{\left| #1 \right\rangle}}

\newcommand{\ie}{{\it{i.e.,~}}}

\newcommand{\comment}[1]{}

\definecolor{gray}{gray}{0.6}

\begin{document}
\title{Optimal conditions for spatial adiabatic passage \\
of a Bose--Einstein condensate}
\author{J.~L.~Rubio}\email[]{juanluis.rubio@uab.cat}
\affiliation{Departament de F\'{\i}sica, Universitat Aut\`{o}noma de Barcelona, E-08193 Bellaterra, Spain} 
\author{V.~Ahufinger}
\affiliation{Departament de F\'{\i}sica, Universitat Aut\`{o}noma de Barcelona, E-08193 Bellaterra, Spain} 
\author{Th.~Busch}
\affiliation{Quantum Systems Unit, Okinawa Institute of Science and Technology Graduate University, Okinawa, 904-0495, Japan}
\author{J.~Mompart}
\affiliation{Departament de F\'{\i}sica, Universitat Aut\`{o}noma de Barcelona, E-08193 Bellaterra, Spain} 

\date{\today}
\begin{abstract} 
We investigate spatial adiabatic passage of a Bose--Einstein condensate in a triple well potential within the three-mode approximation. By rewriting the dynamics in the so-called time-dependent dark/dressed basis, we analytically derive the optimal conditions for the non-linear parameter and the on-site energies of each well to achieve a highly efficient condensate transport. We show that the non-linearity yields a high-efficiency plateau for the condensate transport as a function of the on-site energy difference between the outermost wells, favoring the robustness of the transport. We also analyze the case of different non-linearities in each well, which, for certain parameter values,  leads to an increase of the width of this plateau.

\end{abstract}

%03.65.Xp	Tunneling, traversal time, quantum Zeno dynamics
%03.75.Kk	Dynamic properties of condensates; collective and hydrodynamic excitations, superfluid flow
%03.75.Gg	Entanglement and decoherence in Bose-Einstein condensates
%03.75.Hh	Static properties of condensates; thermodynamical, statistical, and structural properties
%03.75.Kk	Dynamic properties of condensates; collective and hydrodynamic excitations, superfluid flow
%03.75.Lm	Tunneling, Josephson effect, Bose-Einstein condensates in periodic potentials, solitons, vortices, and topological excitations (see also 74.50.+r Tunneling phenomena; Josephson effects in superconductivity)
%03.75.Mn	Multicomponent condensates; spinor condensates
%03.75.Nt	Other Bose-Einstein condensation phenomena
\maketitle

%%%%%%%%%%%%%%%%%%%%%%%%%%%%%%%%%%%%%%%%%%%%%%%%%%%%%%%%%%%%%%%%%%%%%%%%%%%%%%%%%%%%%
\section{Introduction}
\label{sec:introduction}
%%%%%%%%%%%%%%%%%%%%%%%%%%%%%%%%%%%%%%%%%%%%%%%%%%%%%%%%%%%%%%%%%%%%%%%%%%%%%%%%%%%%%
 
Spatial adiabatic passage (SAP) \cite{SAP_review_2016}, i.e., the matter-wave analogue of the well known quantum optical Stimulated Raman Adiabatic Passage (STIRAP) \cite{STIRAP_1,STIRAP_2} technique, has been proposed for  high fidelity and robust transport of quantum particles between the outermost traps of a triple-well potential. SAP is based on adiabatically following an energy eigenstate of the system that, for initial and target resonant wells, consists of a superposition involving only the two localized states of the outermost wells. To transport the quantum particle between these two localized states, the followed eigenstate must be adiabatically transformed from the initial to the target one. SAP processes have been investigated for single-atoms \cite{SAP_Eckert_2004,SAP_Eckert_2006,SAP_Opatrny_2009,SAP_McEndoo_2010,SAP_Morgan_2011,SAP_Benseny_2012,SAP_Morgan_2013,SAP_Menchon_2014a,SAP_Menchon_2014b,SAP_Benseny_2016}, electrons in quantum dots \cite{SAP_Greentree_2004,SAP_Cole_2008,SAP_Huneke_2013}, and experimentally reported for light beams in systems of three evanescently-coupled optical waveguides \cite{Light_SAP_Longhi_2007,Light_SAP_Lahini_2008,Light_SAP_Menchon_2012,Light_SAP_Menchon_2013}.

SAP applied to the transport of a BEC in a triple-well potential has also been investigated \cite{BEC_SAP_Graefe_2006,BEC_SAP_Rab_2008,BEC_SAP_Nesterenko_2009,BEC_SAP_Rab_2012} showing that, in principle, it only works for a very limited range of non-linear interactions \cite{BEC_SAP_Graefe_2006}. In fact, non-linear interactions have two detrimental effects in SAP: (i) the on-site interaction breaks, in a dynamical fashion, the resonant condition between the two outermost wells; and (ii) unstable non-linear eigenstates and level crossings can appear during the dynamics preventing the adiabatic following of an energy eigenstate. 

In this work, we investigate, in the so-called dark/dressed basis \cite{STIRAP_Grigoryan_2009,STIRAP_Boradjiev_2010}, SAP for a BEC in a triple-well potential for arbitrary values of the non-linearity and the energy bias between the wells. In the presence of the non-linearity, we analytically derive the optimal on-site energies of the wells to achieve a highly efficient BEC transport. In particular, we show that the non-linearity relaxes the requirement of degeneracy between the on-site energies of the initial and target wells giving rise to a plateau in the transport efficiency curve as a function of the energy bias between these two wells. The width of this plateau can even be increased by an appropriate local modification of the non-linear BEC parameter.   

The article is organized as follows. Section~\ref{sec:physical_model} introduces the physical model under investigation, which consists of a BEC in a triple-well potential described within the three-mode approximation. The three-mode Hamiltonian of the system is then transformed to the dark/bright and dark/dressed bases in Section~\ref{sec:dark_bright_basis}. Describing the dynamical process in the dark/dressed basis, we analytically derive and numerically check, in Section~\ref{sec:optimal_conditions}, the SAP optimal conditions for the BEC adiabatic transport with an identical non-linear interaction parameter in each well. Section~\ref{sec:Case with multiple factors of non-linearity} is devoted to the discussion of BEC adiabatic transport with different non-linear interaction parameters in each well. Conclusions are presented in Section~\ref{sec:conclusions}.

%%%%%%%%%%%%%%%%%%%%%%%%%%%%%%%%%%%%%%%%%%%%%%%%%%%%%%%%%%%%%%%%%%%%%%%%%%%%%%%%%%%%%
\section{Physical model}
\label{sec:physical_model}
%%%%%%%%%%%%%%%%%%%%%%%%%%%%%%%%%%%%%%%%%%%%%%%%%%%%%%%%%%%%%%%%%%%%%%%%%%%%%%%%%%%%%

\begin{figure}[t!]
\includegraphics[width=0.3\textwidth]{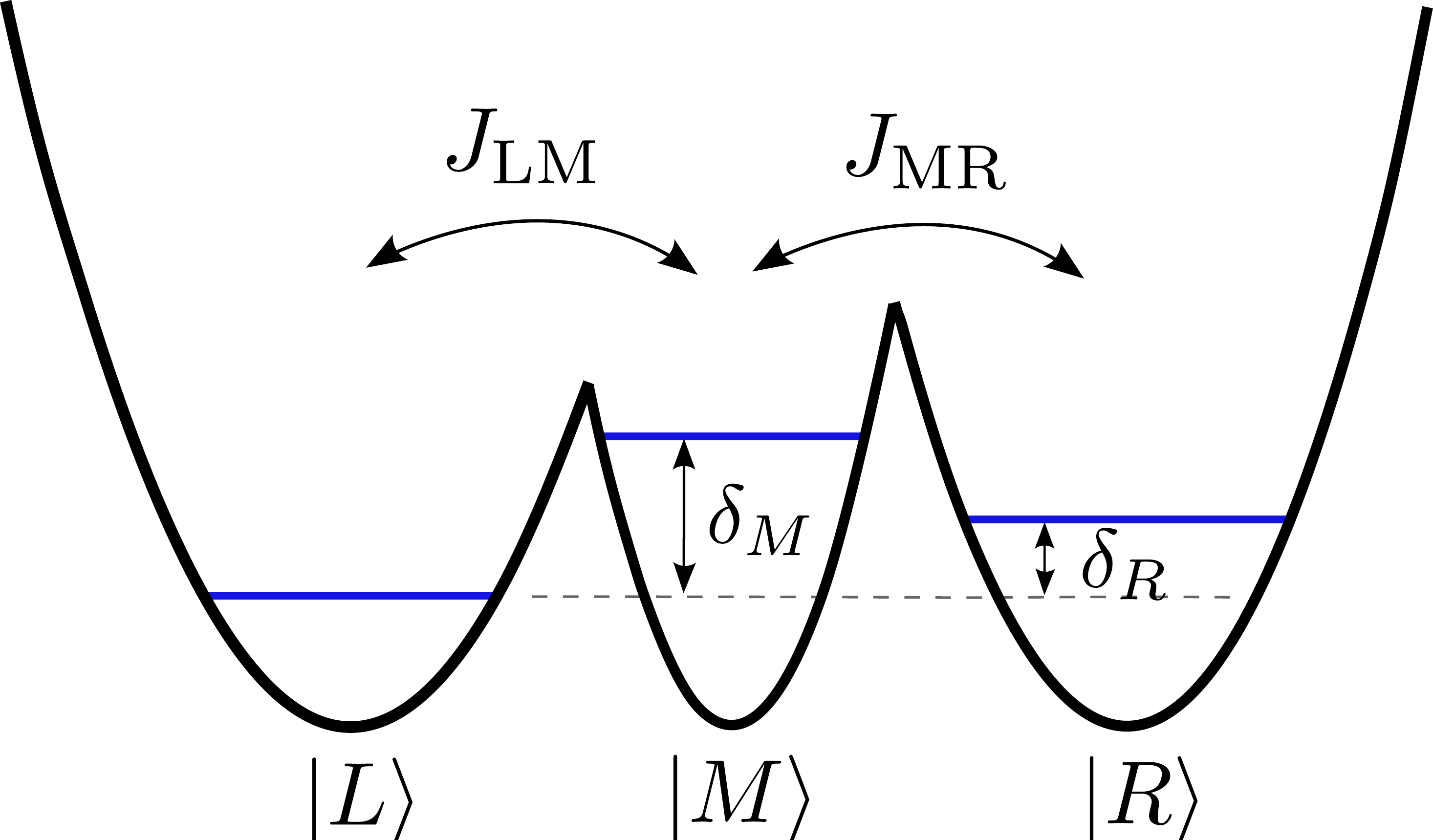}
\caption{
Schematic diagram of the triple-well potential under consideration. $\ket{i}$ with $i=L,M,R$ is the BEC ground state localized in the left, middle, and right wells, respectively, $J_{\rm LM}$ ($J_{\rm MR}$) is the tunneling rate between the left and middle (middle and right) wells, and $\delta_M$ ($\delta_R$) is the  on-site energy of the middle (right) well with respect to the left well.}
\label{fig:1}
\end{figure}

The physical system under investigation consists of a zero temperature BEC in a one-dimensional (1D) triple-well potential with near-neighboring tunneling (see Fig.~\ref{fig:1}).
In the mean-field approximation and at zero temperature, the  evolution of a 1D BEC wave function $\psi(x,t)$ is governed by the Gross--Pitaevskii equation (GPE)
\begin{equation}
\label{GPE}
i\hbar\frac{\partial\psi(x,t)}{\partial t}=H\psi(x,t), \\
\end{equation}
with
\begin{equation}
H=-\frac{\hbar^2}{2m}\frac{\partial^2}{\partial x^2} + V(x,t)+ g_{1D} N |\psi(x,t)|^2.
\end{equation}
Here, $V(x,t)$ is the external trapping potential, $N$ is the total number of atoms in the BEC and $g_{1D}=2 \hbar \omega_{\perp} a_s$ is the 1D non-linear interaction strength being $\omega_{\perp}$ the transverse trapping frequency and $a_s$ the scattering length.

%%%%%%%%%%%%%%%%%%%%%%%%%%%%%%%%%%%%%%%%%%%%%%%%%%%%%%%%%%%%%%%%%%%%%%%%%%%%%%%%%%%%%
%%%%%%%%%%%%%%%%%%%%%%%%%%%%%%%%%%%%%%%%%%%%%%%%%%%%%%%%%%%%%%%%%%%%%%%%%%%%%%%%%%%%%

Assuming that the on-site interaction energy is smaller than the trapping energy of the corresponding well, one can apply the three-mode approximation to describe the BEC dynamics. In this approximation, the BEC wave function or order parameter is written as
% * <juanluis.rubio@gmail.com> 2016-02-25T17:57:30.145Z:
%
% ^
%.
\begin{equation}
\psi(x,t)=a_{L}(t)\phi_{L}(x)+a_{M}(t)\phi_{M}(x)+a_{R}(t)\phi_{R}(x).
\label{three_mode_wavefunction}
\end{equation}
Here $a_{i}$ with $i=L,M,R$ are the probability amplitudes for the BEC to be in the left, middle, and right wells, respectively, while $\phi_i$ is the BEC ground state in the isolated well $i$. We assume $\int_{-\infty}^{+\infty} |\psi (x,t)|^2 dx = 1$ and $\int_{-\infty}^{+\infty} \phi_{j}^{*} \phi_{i} dx =\delta_{ij}$, which means $\sum_i |a_i(t)|^2=1$. Note, however, that for short wells separations, states $\phi_i$ should be properly orthonormalized. From now on, we will use Dirac's notation for the BEC localized states: $\ket{i} \equiv \left\langle x | \phi_{i} \right\rangle $.

\begin{figure}[t!]
\includegraphics[width=0.45\textwidth]{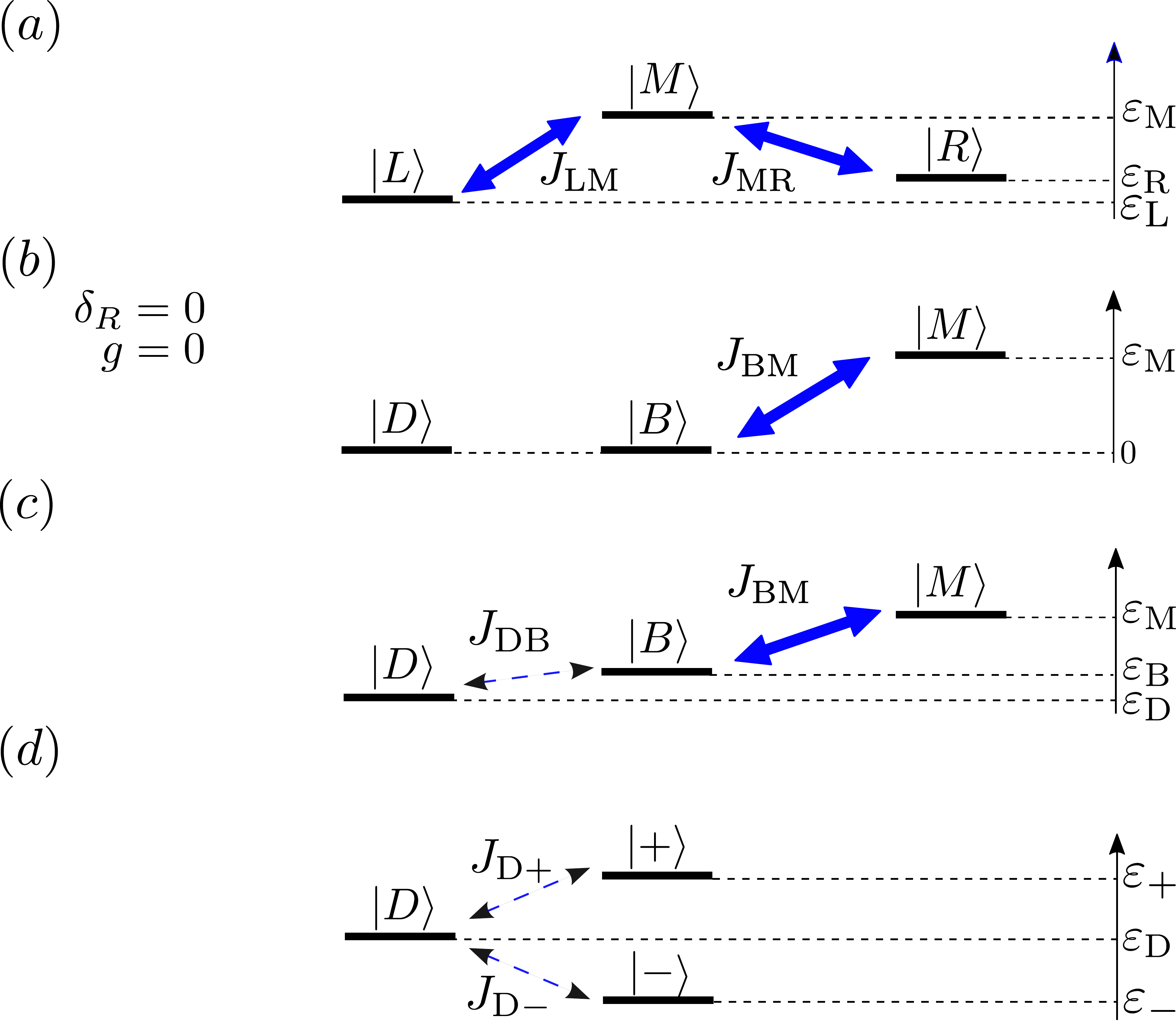}
\caption{(a) Energies and coupling rates in the (a) bare basis $\left\{\ket{L},\ket{M},\ket{R}\right\}$, (b) dark/bright basis $\left\{\ket{D},\ket{B},\ket{M}\right\}$ for $\delta_R=g=0$, (c) dark/bright basis for 
$\delta_R, g \neq 0$, and (d) dark/dressed basis $\left\{\ket{D},\ket{+},\ket{-}\right\}$ for 
$\delta_R, g \neq 0$.}
\label{fig:2}
\end{figure}

Under the three-mode approximation, the equations of motion for the probability amplitudes $a_i$ can be derived casting \eqref{three_mode_wavefunction} in $\eqref{GPE}$ yielding
\begin{equation}
\label{GPEH1}
	i \hbar \frac{d}{dt}
		\begin{pmatrix}
		a_{L}\\
		a_{M}\\
		a_{R}
		\end{pmatrix} 	 
 =	\hat{H}_{LMR}
		\begin{pmatrix}
		a_{L}\\
		a_{M}\\
		a_{R}
		\end{pmatrix},
\end{equation}
with the Hamiltonian \cite{BEC_SAP_Graefe_2006}
\begin{equation}
\label{bare_hamiltonian}
	  \hat{H}_{LMR}= \hbar		
		\begin{pmatrix}
		g_{L} |a_L|^2 					& -\frac{J_{\rm LM}}{2}             & 0                       \\
		-\frac{J_{\rm LM}}{2}   & \delta_M + g_{M} |a_M|^2 		& -\frac{J_{\rm MR}}{2}       \\
		0         					& -\frac{J_{\rm MR}}{2}             & \delta_R + g_{R} |a_R|^2	
		\end{pmatrix},
\end{equation}
where $\varepsilon_{i}\equiv\delta_i+g_{i} |a_{i}|^2$ are the energies in the bare basis $\left\{\ket{L},\ket{M},\ket{R}\right\}$, being $\delta_L=0$ for the left well. The on-site single-atom energy bias for middle ($\delta_M$) and right ($\delta_R$) wells with respect to the left one read
\begin{equation}
\label{defdetunings}
\hbar \delta_{i} = \int_{-\infty}^{+\infty} \phi_{i}^* \left(  -\frac{\hbar^2}{2m} \frac{\partial^2}{\partial x^2} + V(x,t) \right)\phi_{i}\,dx ,
\end{equation}
the tunneling rates between the left and middle, and between middle and right wells are
\begin{subequations}
\label{defcouplings}
\begin{eqnarray}
- \frac{\hbar}{2} J_{\rm LM} &=& \int_{-\infty}^{+\infty} \phi_M^* H_{LMR}\,\phi_{L} dx,\;\;\;\;\;\;\;\; \\
- \frac{\hbar}{2} J_{\rm MR} &=& \int_{-\infty}^{+\infty} \phi_R^* H_{LMR}\,\phi_{M} dx,\;\;\;\;\;\;\;\;
\end{eqnarray}
\end{subequations}
respectively, and the non-linear interaction parameter are
\begin{equation}
\label{defg}
\hbar g_{i} = g_{1D} N \int_{-\infty}^{+\infty} |\phi_i (x)|^4 dx.
\end{equation}
for each well $i$, which, in Sections \ref{sec:dark_bright_basis} and \ref{sec:optimal_conditions}, have been assumed to be the same and equal to $g$ for the three wells, while in Section \ref{sec:Case with multiple factors of non-linearity} we have considered the case of a different non-linear interaction parameter in each well. Energies and coupling rates in the bare basis are sketched in Fig.~\ref{fig:2}(a).

%%%%%%%%%%%%%%%%%%%%%%%%%%%%%%%%%%%%%%%%%%%%%%%%%%%%%%%%%%%%%%%%%%%%%%%%%%%%%%%%%%%%%
%%%%%%%%%%%%%%%%%%%%%%%%%%%%%%%%%%%%%%%%%%%%%%%%%%%%%%%%%%%%%%%%%%%%%%%%%%%%%%%%%%%%%

%%%%%%%%%%%%%%%%%%%%%%%%%%%%%%%%%%%%%%%%%%%%%%%%%%%%%%%%%%%%%%%%%%%%%%%%%%%%%%%%%%%%%
\section{SAP in the dark/bright and dark/dressed bases}
\label{sec:dark_bright_basis}
%%%%%%%%%%%%%%%%%%%%%%%%%%%%%%%%%%%%%%%%%%%%%%%%%%%%%%%%%%%%%%%%%%%%%%%%%%%%%%%%%%%%%

For $g_{i}=0$, \ie for the single particle case, and $\delta_R=0$, the diagonalization of the three-mode Hamiltonian (\ref{bare_hamiltonian}) gives three energy eigenstates. One of them is the so-called spatial dark state 
\begin{equation}
\ket{D}=\cos \theta \ket{L} - \sin \theta \ket{R}
\label{dark_state}
\end{equation} 
with $\tan\theta=J_{\rm LM}/J_{\rm MR}$. SAP consists of adiabatically following the spatial dark state (\ref{dark_state}) by slowly varying the mixing angle $\theta$ from $0 $ to $\pi/2$ to transport a quantum particle from $\ket{L}$ to $\ket{R}$.

To investigate the transport of a BEC from the left to the right well via SAP we proceed in two steps. First, we transform Hamiltonian (\ref{bare_hamiltonian}) to the dark/bright basis and, later on, to the dark/dressed basis. The latter allows to derive the optimal conditions for the transport process.

\subsection{Dark/bright basis}

The dark state given by Eq.~(\ref{dark_state}), the bright state $\ket{B}=\sin\theta\ket{L}+\cos\theta\ket{R}$ fulfilling $\left\langle D | B \right\rangle=0$, and the $\ket{M}$ state define the dark/bright basis. Since dark and bright states are time dependent through $\theta(t)$, the transformed Hamiltonian is given by
\begin{equation}
\label{Hchange}
\hat{H}_{DBM}=C\hat{H}_{LMR}C^{-1}+i\hbar\frac{d C}{dt}C^{-1},
\end{equation}
where
\begin{equation}
\label{C1}
C=\begin{pmatrix}
		\cos\theta &  0  &  -\sin\theta  \\
		\sin\theta &  0	 &  \cos\theta   \\
		0	&  1  & 0	
		\end{pmatrix} .
\end{equation}
Assuming a successful SAP process, i.e., $|a_L|^2=\cos^2 \theta $, $|a_M|^2=0$, and $|a_R|^2=\sin^2 \theta$, and casting $\eqref{bare_hamiltonian}$ and $\eqref{C1}$ in $\eqref{Hchange}$ we obtain
\begin{equation}
\label{dark_bright_hamiltonian}
	  \hat{H}_{DBM}= \hbar		
		\begin{pmatrix}
		\varepsilon_D 	  &  -\frac{J_{\rm DB}}{2}-i\dot{\theta}   &  -\frac{J_{\rm DM}}{2}     \\
		-\frac{J_{\rm DB}}{2}+i\dot{\theta} &  \varepsilon_B		   &  -\frac{J_{\rm BM}}{2}     \\
		-\frac{J_{\rm DM}}{2}	&  -\frac{J_{\rm BM}}{2}   & \varepsilon_M	
		\end{pmatrix},
\end{equation}
with
\begin{eqnarray}
\varepsilon_D &=& \frac{1}{4}\left(g\left(3+\cos4\theta\right)+2\delta_R\left(1-\cos2\theta\right)\right),\label{ED} \\
\varepsilon_{B} &=& \cos^2\theta\left(\delta_{R}+2g\sin^2\theta\right),\label{EB} \\
\varepsilon_{M} &=& \delta_{M}, \label{EM} \\
J_{\rm DB} &=& \left(\delta_R-g\cos2\theta\right)\sin 2\theta, \label{JDB} \\
J_{\rm DM} &=& J_{\rm LM}\cos\theta-J_{\rm MR}\sin\theta=0, \label{JDM} \\ 
J_{\rm BM} &=& J_{\rm LM}\sin\theta+J_{\rm MR}\cos\theta=\sqrt{J_{\rm LM}^2+J_{\rm MR}^2}, \label{JBM}
\end{eqnarray}
where the definition of the mixing angle has been used in $\eqref{JDM}$ and $\eqref{JBM}$, and $g=g_{L}=g_{M}=g_{R}$. Note that $\varepsilon_{D}=\varepsilon_{B}=0$ for $\delta_{R}=g=0$,  {corresponding to the case displayed in Fig.~\ref{fig:2}(b). For the general case sketched in Fig.~\ref{fig:2}(c), it is straightforward to check that $J_{\rm BM}\gg J_{\rm DB}$. 

\subsection{Dark/dressed basis}

Let us proceed now by diagonalizing the $2\times 2$ submatrix associated to states \ket{B} and \ket{M} in Hamiltonian (\ref{dark_bright_hamiltonian}), obtaining the dressed states
\begin{equation}
\label{pm}
\ket{\pm}=\cfrac{1}{N_{\pm}}\left(\ket{M}+\zeta_{\pm}\ket{B}\right),
\end{equation}
where
\begin{eqnarray}
\label{zeta}
\zeta_{\pm} &\equiv& \cfrac{1}{J_{\rm BM}}\left(\varepsilon_{M}-\varepsilon_{B}\pm\sqrt{\left(\varepsilon_{M}-\varepsilon_{B}\right)^2+J_{\rm BM}^2}\right),\;\;\;\; \\
\label{N}
N_{\pm} &\equiv& \sqrt{\zeta_{\pm}^2+1}.\;\;\;\;
\end{eqnarray}
with the corresponding energies
\begin{eqnarray}
\label{dressed_energies}
\varepsilon_{\pm} &=& \frac{1}{8}\left(\xi+4\delta_M\pm\sqrt{\left(\xi-4\delta_M\right)^2+16 J_{\rm BM}^2}\right),
\end{eqnarray}
where
\begin{equation}
\label{xi}
\xi \equiv g\left(1-\cos4\theta\right)+2\delta_R\left(1+\cos2\theta\right).
\end{equation}

Since $\ket{D}$ and $\ket{\pm}$ are time-dependent, proceeding as in the previous subsection, the Hamiltonian in the $\left\{\ket{D},\ket{+},\ket{-}\right\}$ basis transformed from $H_{LMR}$ reads
\begin{eqnarray}
\label{dark_dressed_hamiltonian}
	  \hat{H}_{D\pm}&=&\hbar		
		\begin{pmatrix}
		\varepsilon_{D}	& -\frac{J_{\rm D+}}{2} & -\frac{J_{\rm D-}}{2} \\
		-\frac{J_{\rm D+}}{2} & \varepsilon_{-} & 0 \\
		-\frac{J_{\rm D-}}{2} & 0 & \varepsilon_{+}		
		\end{pmatrix} \nonumber \\
		&+& 
		i\hbar \dot{\theta} \begin{pmatrix}
		0	& -\frac{\zeta_{+}}{N_{+}} & -\frac{\zeta_{-}}{N_{-}} \\
		\frac{\zeta_{+}}{N_{+}} & 0 & -\frac{\dot{\zeta_{+}}}{\zeta_{+}N_{+}^{2}} \\
		\frac{\zeta_{-}}{N_{-}} & \frac{\dot{\zeta_{+}}}{\zeta_{+}N_{+}^{2}} & 0		
		\end{pmatrix},
\end{eqnarray}
where
\begin{equation}
J_{\rm D\pm} =\cfrac{\zeta_{\pm}}{N_{\pm}}J_{\rm DB}\\
\end{equation}
are the couplings between the dark and the dressed states. 
Adiabaticity of the process \cite{STIRAP_2}, $\dot{\theta}\rightarrow 0$, allows to  neglect the second matrix at the r.h.s. of (\ref{dark_dressed_hamiltonian}) since all elements of this matrix are finite as $\dot{\zeta_{\pm}}\propto\zeta_{\pm}^{2}\dot{\theta}$.

Note that, for $g=\delta_R=0$, the expressions for the involved energies, (\ref{ED}) and (\ref{dressed_energies}), take the form
\begin{eqnarray}
\label{energies}
\varepsilon_D &=& 0, \\
\varepsilon_{\pm} &=& \frac{1}{2}\left(\delta_M\pm\sqrt{\delta_M^2+J_{\rm BM}^2}\right).
\end{eqnarray}
This last expression gives the well-known energy shifts in the absence of non-linear interaction.

In the presence of the non-linearity, as the spatial dark state to be followed is weakly coupled to the dressed states $\ket{\pm}$ (see Fig.~\ref{fig:2}(d)), its adiabatic following will fail whenever the dark state gets on resonance with any of the two dressed states. Thus, the necessary condition for SAP to work is
\begin{equation}
\label{TZ condition}
\varepsilon_- (t) < \varepsilon_D (t) < \varepsilon_+ (t),\;\;\;\;\forall t\in[t_i,t_f],
\end{equation} 
where $t_i$ and $t_f$ are the initial and final times of the process, respectively. Hereinafter, we will refer to the set of parameter values satisfying $\eqref{TZ condition}$ as the \textit{Optimal Zone} (OZ).

As mentioned above, the SAP protocol consists of adiabatically following the dark state $\eqref{dark_state}$ by varying the mixing angle from 0 to $\pi/2$. Figs.~$\ref{fig:3}$(a)-(c)
show $\varepsilon_D$ and $\varepsilon_{\pm}$ given in (\ref{ED}) and (\ref{dressed_energies}), respectively, for the following temporal variation of the couplings (see Fig.~$\ref{fig:3}$(d))
\begin{eqnarray}
\label{couplings}
J_{\rm LM} &=& J_0 e^{-\frac{(t-t_p)^2}{2\sigma^2}}, \\
J_{\rm MR} &=& J_0 e^{-\frac{(t-t_s)^2}{2\sigma^2}},
\end{eqnarray}
with $t_p=T/2$, $t_s=-T/2$, $T=1.5\sigma$, $\sigma=150$, and $J_0=1$, from $t_i=-600$ to $t_f=600$, all in harmonic oscillator (h.o.) units with respect to the trapping frequency, $\omega_{x}$, of each well.  

In the absence of the non-linearity, $g=0$, and with the three localized states being on resonance, \ie $\delta_M=\delta_R=0$, the energies are $\varepsilon_{D}=0$, and $\varepsilon_{\pm}(t)=\pm J_{\rm BM}(t)/2$, see Fig.~$\ref{fig:3}$(a), which means that condition $\eqref{TZ condition}$ is fulfilled for all times.
Fig.~$\ref{fig:3}$(b) shows the energies $\varepsilon_{D}$ and $\varepsilon_{\pm}$ for $g\neq0$ and $\delta_R=0$.
The presence of $g$ shifts the energy of the dark state which, in principle, could cross any of the two dressed states. However, by  appropriately selecting the value of $\delta_{M}$, one can compensate this shift and bring back $\varepsilon_D$ to the OZ. This occurs for $\delta_{M}>g$, as shown in Fig.~$\ref{fig:3}$(b) for $g=0.1$ and $\delta_{M}=0.15$.
For $\delta_{R}\not=0$, see Fig.~$\ref{fig:3}$(c), $\varepsilon_D$ is asymmetrically shifted at both temporal extremes of the process, and $\varepsilon_{\pm}$ are also modified depending on the relative values between $\delta_M$ and $\delta_R$.
In the figure, $\varepsilon_D$ crosses $\varepsilon_{+}$, despite the fact that the process starts within the OZ. In this scenario, we have a resonance between the dark state \ket{D} and the dressed state \ket{+} that will break the adiabaticity of the process. In Fig.~$\ref{fig:3}$(c), we have used the values $g=0.1$, $\delta_{M}=0.05$, and $\delta_{R}=0.3$.

%%%%%%%%%%%%%%%%%%%%%%%%%%%%%%%%%%%%%%%%%%%%%%%%%%%%%%%%%%%%%%%%%%%%%%%%%%%%%%%%%%%%%
\begin{figure}[t]
\includegraphics[width=0.5\textwidth]{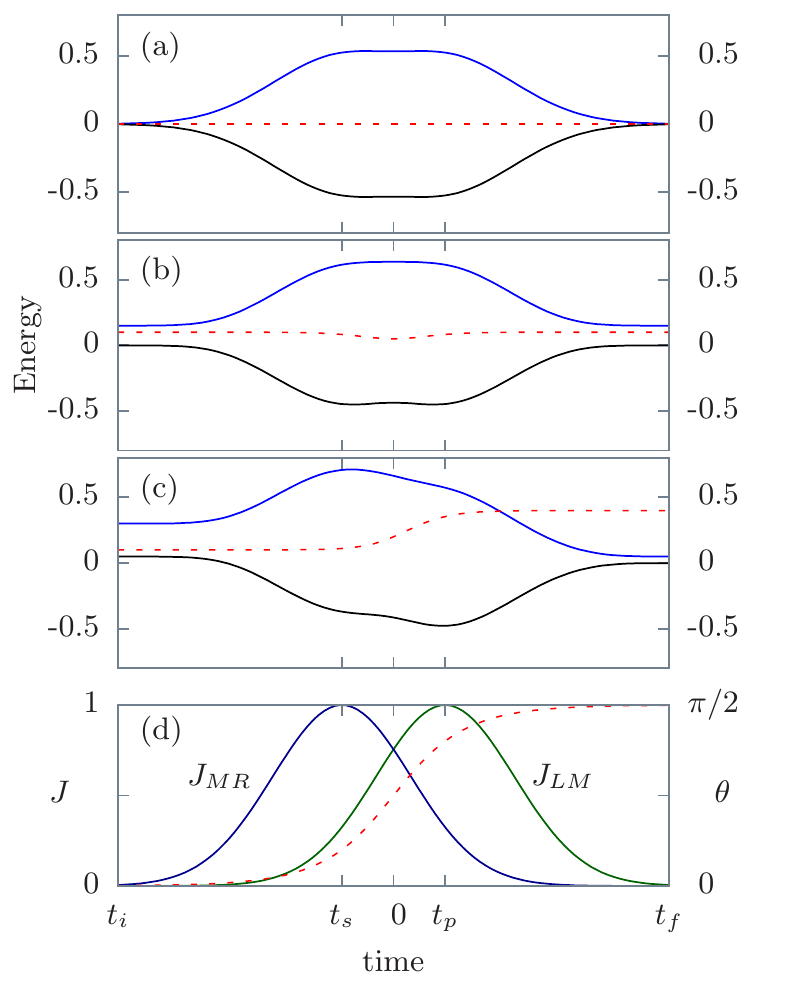}
\caption{ 
Energies of the dark and the dressed states as a function of time: $\varepsilon_{D}$ (red dashed), $\varepsilon_{+}$ (blue solid), and $\varepsilon_{-}$ (black solid) for (a) $g=\delta_{M}=\delta_{R}=0$, (b) $g=0.1$, $\delta_{M}=0.15$, $\delta_{R}=0$, and (c) $g=0.1$, $\delta_{M}=0.05$, $\delta_{R}=0.3$. (d) Temporal sequence of the tunneling rates (solid color lines) and of the mixing angle (dashed red line). Parameters and variables in h.o. units. See text for more details.}
\label{fig:3}
\end{figure}
%%%%%%%%%%%%%%%%%%%%%%%%%%%%%%%%%%%%%%%%%%%%%%%%%%%%%%%%%%%%%%%%%%%%%%%%%%%%%%%%%%%%%

In what follows, we will study condition $\eqref{TZ condition}$ in more detail.

%%%%%%%%%%%%%%%%%%%%%%%%%%%%%%%%%%%%%%%%%%%%%%%%%%%%%%%%%%%%%%%%%%%%%%%%%%%%%%%%%%%%%
\section{SAP optimal conditions for a BEC}
\label{sec:optimal_conditions}
%%%%%%%%%%%%%%%%%%%%%%%%%%%%%%%%%%%%%%%%%%%%%%%%%%%%%%%%%%%%%%%%%%%%%%%%%%%%%%%%%%%%%

In this section, we will show that conditions
\begin{subequations}
\label{TZ condition_tif}
\begin{eqnarray}
\label{TZ condition_ti}
\varepsilon_- (t_{i}) < \varepsilon_D (t_{i}) < \varepsilon_+ (t_{i}), \\
\label{TZ condition_tf}
\varepsilon_- (t_{f}) < \varepsilon_D (t_{f}) < \varepsilon_+ (t_{f}),
\end{eqnarray}
\end{subequations}
imply the most general condition $\eqref{TZ condition}$ if $J_{0}$ is larger than a threshold value $J_{0,\rm{min}}$. The initial, $t_{i}$, and final, $t_{f}$, times are defined such that the tunneling rates $J_{\rm LM}(t_{i,f}),J_{\rm MR}(t_{i,f})\rightarrow 0$, with  $J_{\rm MR}(t_i)>J_{\rm LM}(t_i)$ and $J_{\rm LM}(t_f)>J_{\rm MR}(t_f)$. Therefore, $\theta(t_i)\approx 0$ and $\theta(t_f)\approx\pi/2$.

Let us consider the optimal zones OZ$(t_i)$ and OZ$(t_f)$ at the initial and final time of the process. Imposing $\varepsilon_{\pm}(t_{i})=\varepsilon_{D}(t_{i})$ and $\varepsilon_{\pm}(t_{f})=\varepsilon_{D}(t_{f})$, we obtain the pair of curves ${\rm Ci}$ and ${\rm Cf}$:
\begin{subequations}
\label{CiCf}
\begin{eqnarray}
{\rm Ci}\rightarrow\delta_{R} &=& g+\frac{J_{\rm MR}^2(t_i)}{2(\delta_M-g)},\;\;\\
{\rm Cf}\rightarrow\delta_{R} &=& -g+\frac{1}{2}\left(\delta_M\pm\sqrt{J_{\rm LM}^2(t_f)+\delta_M^2}\right),\;\;
\end{eqnarray}
\end{subequations}
which fix an interval of parameters allowing $\varepsilon_D$ to remain in the OZ at times $t_i$ and $t_f$, respectively.
Curves \eqref{CiCf} allow to find the boundaries of the intersection region OZ$(t_i)\cap$OZ$(t_f)$. Inside this region, the set of parameters $\{g, \delta_{M}, \delta_{R}\}$ fulfill
\begin{subequations}
\label{TZ inequality0+}
\begin{eqnarray}
\label{TZ inequality1+}
\delta_{M}>g, \\
\label{TZ inequality2+}
1>\cfrac{\delta_{R}}{g}>-1, \\
\label{TZ inequality3+}
\delta_M>\delta_{R}+g,
\end{eqnarray}
\end{subequations}
if $g>0$, and 
\begin{subequations}
\label{TZ inequality0-}
\begin{eqnarray}
\label{TZ inequality1-}
\delta_{M}<g, \\
\label{TZ inequality2-}
1<\cfrac{\delta_{R}}{g}<-1, \\
\label{TZ inequality3-}
\delta_M<\delta_{R}+g,
\end{eqnarray}
\end{subequations}
if $g<0$. Note that, for $g>0$ and $\delta_{R}=0$, we can keep the system in OZ$(t_i)\cap$OZ$(t_f)$ simply fulfilling the first condition $\eqref{TZ inequality1+}$, which is consistent with the conclusions of Ref. \cite{BEC_SAP_Graefe_2006}. Inequalities $\eqref{TZ inequality0+}$ for $g>0$ ($\eqref{TZ inequality0-}$ for $g<0$) determine the set of parameters that keeps the dark state out of resonance with the dressed states at  $t_i$ and $t_f$, fulfilling $\eqref{TZ condition_tif}$.
 
Once we have explicitly obtained the above conditions, it is possible to show that $\eqref{TZ condition_tif}$ implies the most general condition $\eqref{TZ condition}$, \ie OZ=OZ$(t_i)\cap$OZ$(t_f)$.
With this aim, it is convenient to consider the involved energies as a function of $\theta$ within the range $\theta\in (\theta_{i},\theta_{f})$, where $\theta_{i}=\theta(t=t_{i})=0$, and $\theta_{f}=\theta(t=t_{f})=\pi/2$. 
In Appendix~A we demonstrate that 
$\eqref{TZ inequality0+}$-$\eqref{TZ inequality0-}$ imply $\varepsilon_{-}(\theta)<\varepsilon_{D}(\theta)<\varepsilon_{+}(\theta)$ for $0<\theta<\pi/2$.
To do this, we just consider the energy curves and their extreme values. While the crossing between $\varepsilon_{D}(\theta)$ and $\varepsilon_{+}(\theta)$ ($\varepsilon_{-}(\theta)$) is avoided for $g>0$ $(g<0)$, it exists, however, a minimum value of the tunneling rate given by
\begin{equation}
\label{J0}
J_{0,\rm{min}}=\cfrac{1}{2}\left|\cfrac{\delta_{R}}{g}\right|\sqrt{\cfrac{\delta_{R}^2}{2}+4g\delta_{M}-2g\delta_{R}-2g^2}
\end{equation}
above which $\varepsilon_{D}(\theta)$ and $\varepsilon_{-}(\theta)$ ($\varepsilon_{+}(\theta)$) do not intersect for $g>0$ $(g<0)$. Note that $J_{0,\rm{min}}=0$ for $\delta_{R}=0$, while it takes a positive real value for arbitrary values of $\delta_R$, $\delta_M$, and $g$ fulfilling $\eqref{TZ inequality0+}$ or $\eqref{TZ inequality0-}$. Thus, we conclude that the problem of finding the optimal conditions to efficiently perform SAP of a BEC is reduced to fulfill conditions $\eqref{TZ inequality0+}$-$\eqref{TZ inequality0-}$, which are derived at the initial and final times of the process, together with $J_{0} > J_{0,\rm{min}}$.

%%%%%%%%%%%%%%%%%%%%%%%%%%%%%%%%%%%%%%%%%%%%%%%%%%%%%%%%%%%%%%%%%%%%%%%%%%%%%%%%%%%%%
\begin{figure}[t]
\includegraphics[width=0.45\textwidth]{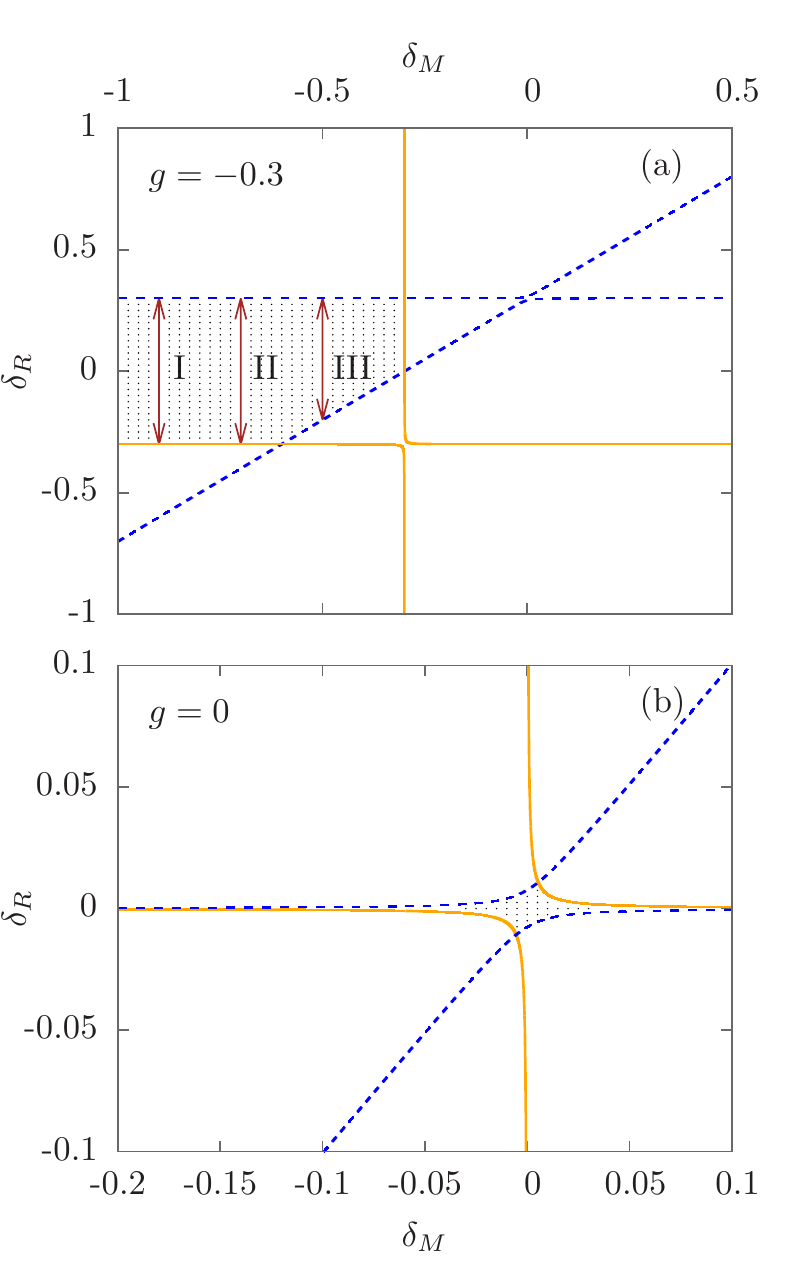}
\caption{
OZ (dotted pattern) in the parameter plane $(\delta_{M},\delta_{R})$ 
for the SAP of a BEC from \ket{L} to \ket{R} with (a) $g=-0.3$ and (b) $g=0$, delimited by the curves ${\rm Ci}$ (orange solid lines) and ${\rm Cf}$ (blue dashed lines). In (a), red vertical arrows I, II and III indicate the optimal range of $\delta_R$ values for $\delta_M$ equal to $-0.9,-0.7$ and $-0.5$, respectively. Note that in (b) the OZ is much smaller than in (a). The temporal variation of the tunnelings is the same as in Fig.~\ref{fig:3}(d). Parameters and variables in h.o. units.
}
\label{fig:4}
\end{figure}
%%%%%%%%%%%%%%%%%%%%%%%%%%%%%%%%%%%%%%%%%%%%%%%%%%%%%%%%%%%%%%%%%%%%%%%%%%%%%%%%%%%%%

%%%%%%%%%%%%%%%%%%%%%%%%%%%%%%%%%%%%%%%%%%%%%%%%%%%%%%%%%%%%%%%%%%%%%%%%%%%%%%%%%%%%%
\begin{figure*}[t!]
\centering
\includegraphics[width=1\textwidth]{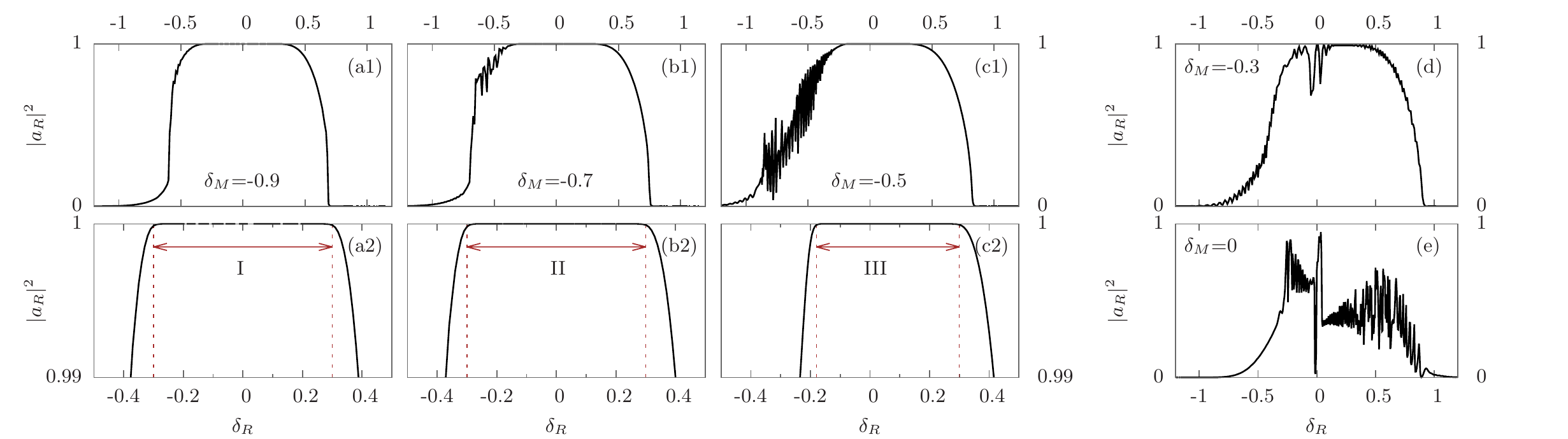}
\caption{
Numerically calculated transfer efficiency for the SAP of a BEC from state \ket{L} to \ket{R} with $g=-0.3$ as a function of $\delta_R$ for (a1,a2) $\delta_M=-0.9$, (b1,b2) $\delta_{M}=-0.7$, (c1,c2) $\delta_{M}=-0.5$, (d) $\delta_{M}=-0.3$, and (e) $\delta_{M}=0$. Red horizontal arrows in (a2), (b2), and (c2) indicate the width of the plateaus for the maximum efficiency, in excellent agreement with the predicted OZs indicated in Fig.~\ref{fig:4} with red vertical arrows. 
The temporal variation of the tunnelings is the same as in Fig.~\ref{fig:3}(d).
Parameters and variables in h.o. units.
}
\label{fig:5}
\end{figure*}
%%%%%%%%%%%%%%%%%%%%%%%%%%%%%%%%%%%%%%%%%%%%%%%%%%%%%%%%%%%%%%%%%%%%%%%%%%%%%%%%%%%%%

%
%%%%%%%%%%%%%%%%%%%%%%%%%%%%%%%%%%%%%%%%%%%%%%%%%%%%%%%%%%%%%%%%%%%%%%%%%%%%%%%%%%%%%
%\section{NUMERICAL SIMULATIONS}
%\label{sec:NUMERICALSIMULATIONS}
%%%%%%%%%%%%%%%%%%%%%%%%%%%%%%%%%%%%%%%%%%%%%%%%%%%%%%%%%%%%%%%%%%%%%%%%%%%%%%%%%%%%%

Fig.~$\ref{fig:4}$ shows curves $\rm Ci$ (orange solid lines) and $\rm Cf$ (blue dashed lines) defined in (\ref{CiCf}), for $g=-0.3$ (Fig.~$\ref{fig:4}$(a)) and $g=0$ (Fig.~$\ref{fig:4}$(b)), with the same temporal variation for the couplings used in Fig.~$\ref{fig:3}$(d). For $g=-0.3$, the resulting OZ has been highlighted with a dotted pattern. Note that for $\delta_M < -0.6$, the OZ extends symmetrically in the range $\delta_R \in (-g,g)$. However, for $g=0$, the resulting OZ reduces to a very small region} around $\delta_R=\delta_M=0$. Thus, we conclude that the presence of the non-linearity broadens the energy bias range for which the adiabatic transport succeeds.

By means of the numerical integration of Eq.~$\eqref{GPEH1}$, 
Fig.~$\ref{fig:5}$ shows the SAP efficiency as a function of $\delta_R$ for different values of $\delta_{M}$ and the rest of parameter values as in Fig.~$\ref{fig:4}$(a). Note that the numerically obtained plateaus of maximum efficiency in Figs.~$\ref{fig:5}$(a), (b) and (c), highlighted with red horizontal arrows, correspond to the values of $\delta_{R}$ predicted in Fig.~$\ref{fig:4}$(a): plateaus I and II range in the interval $\delta_{R}\in\left(-0.3,0.3\right)$ for $\delta_{M}=-0.9$ (Fig.~$\ref{fig:5}$(a2)) and $\delta_{M}=-0.7$ (Fig.~$\ref{fig:5}$(b2)), and the one labeled with III extends along the interval $\delta_{R}\in\left(-0.2,0.3\right)$ for $\delta_{M}=-0.5$ (Fig.~$\ref{fig:5}$(c2)).
%
%%%%%%%%%%%%%%%%%%%%%%%%%%%%%%%%%%%%%%%%%%%%%%%%%%%%%%%%%%%%%%%%%%%%%%%%%%%%%%%%%%%%%
\begin{figure}[t!]
\centering
\includegraphics[width=0.45\textwidth]{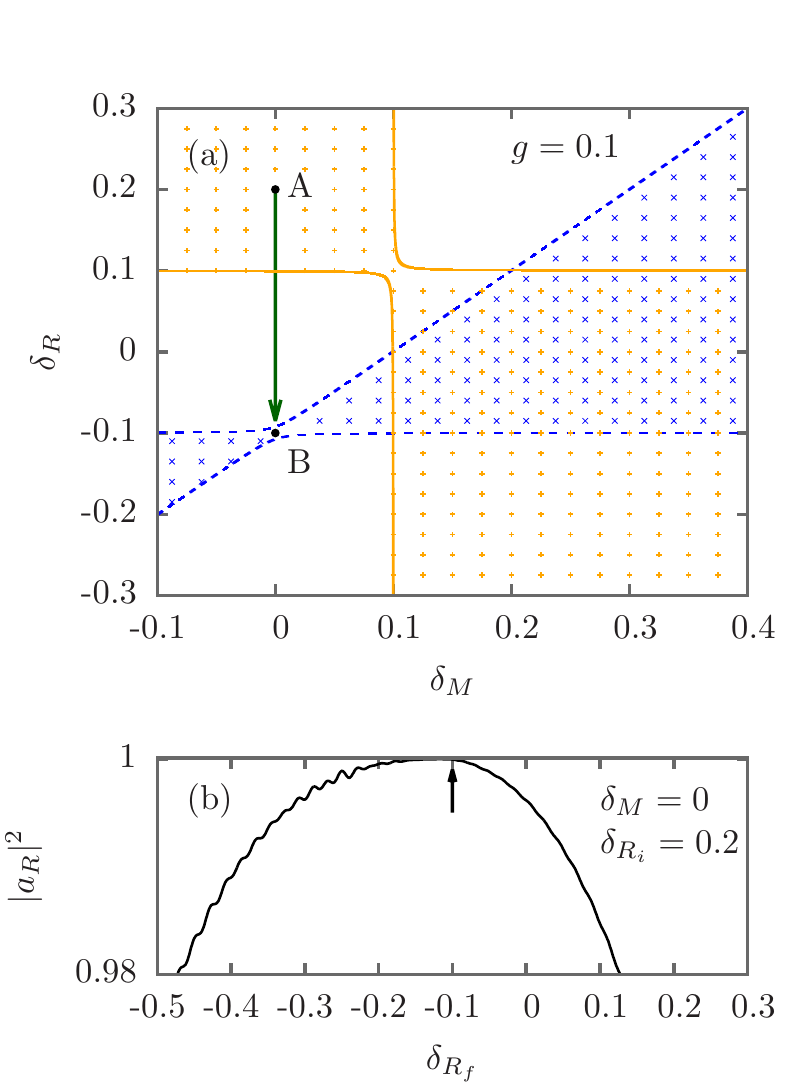}
\caption{(a) OZ$(t_i)$ (orange crosses pattern) and OZ$(t_f)$ (blue blades pattern) delimited by the curves ${\rm Ci}$ (orange solid lines) and ${\rm Cf}$ (blue dashed lines), respectively, for $g=0.1$. A and B correspond to $\left(\delta_{R_{i}},\delta_{M}\right)$ and $\left(\delta_{R_{f}},\delta_{M}\right)$ belonging to OZ$(t_i)$ and OZ$(t_f)$, respectively. (b) Efficiency for the SAP of a BEC as a function of $\delta_{R_{f}}$ for $\delta_{R_{i}}=0.2$ (point A) and using a linear dependence in time for $\delta_{R}(t)$ along the process. The maximum transfer probability (indicated with a black arrow) corresponds to the value $\delta_{R_{f}}=-0.1$ (point B).
The temporal variation of the tunnelings is the same as in Fig.~\ref{fig:3}(d). Parameters and variables in h.o. units.
}
\label{fig:6}
\end{figure}
%%%%%%%%%%%%%%%%%%%%%%%%%%%%%%%%%%%%%%%%%%%%%%%%%%%%%%%%%%%%%%%%%%%%%%%%%%%%%%%%%%%%%
%
For $\delta_M=-0.3$ (Fig.~$\ref{fig:5}$(d)), although there is a broad region with efficiency values close to 1, fluctuations appear. For the simulated value outside the OZ, $\delta_M=0$ (Fig.~$\ref{fig:5}$(e)), fluctuations dramatically increase, and adiabatic transport fails.

Up to now we have analyzed the optimal conditions for the SAP of a BEC, summarized in $\eqref{TZ inequality0+}$ and $\eqref{TZ inequality0-}$, by using constant values of the parameters.
Nonetheless, we can also dynamically change the value of some parameters during the process, as long as our system remains within the OZ for all times $t \in (t_{i},t_{f})$. For instance, for a fixed value of $g$ and a value of $\delta_M$ in OZ$(t_i)$, we can appropriately vary $\delta_R$ in time from $\delta_{R_{i}}=\delta_{R}(t_{i})$ to $\delta_{R_{f}}=\delta_{R}(t_{f})$, as schematically shown in Fig.~$\ref{fig:6}$(a), keeping the pair $\left(\delta_{M},\delta_{R}\left(t\right)\right)\in$ OZ$(t)\,$ $\forall$t.
In Fig.~$\ref{fig:6}$(b), the numerically calculated efficiency for the SAP of a BEC with $g=0.1$ by fixing the value $\delta_{M}=0$ and starting from $\delta_{R_{i}}=0.2$ (point A) is plotted as a function of the final value $\delta_{R_{f}}$, showing that the maximum efficiency is obtained when the process ends at point B, which corresponds to a value $\delta_{R}=-0.1$ belonging to the OZ$(t_{f})$. For the simulation we have used a linear dependence $\delta_{R}(t)=m\,t+n$ with $m=\left(\delta_{R_{i}}-\delta_{R_{f}}\right)/\left(t_{i}-t_{f}\right)$, and $n=\delta_{R_{f}}-m\,t_{f}$. 

Note, in addition, that an alternative protocol to achieve a high efficient SAP of a BEC from the left to the right well consists of considering a temporal variation of the right well energy bias fulfilling $\delta_R (t) = g \cos 2\theta (t)$ such that $J_{DB}=0$, see Eq.~\eqref{JDB}. This is exactly the approach suggested in \cite{SAP_Morgan_2011}.

To conclude this section, we will make an estimation of the applicability of the SAP protocol to state-of-the-art BECs. Applying Eq.~\eqref{defg} to a BEC composed of 10000 $^{87}$Rb atoms,  with $s$-wave scattering length $a_s=5.3\,{\rm nm}$, trapped in a potential whose transverse trapping frequency is $\omega_{\perp}=2\pi \times 1000\,{\rm Hz}$, one obtains $g=0.5$ in h.o. units. As predicted by inequality \eqref{TZ inequality1+}, SAP of this $^{87}$Rb BEC is not possible at full resonance, i.e., for $\delta_{R}=\delta_{M}=0$, but it will work for $\delta_{M}>g$. At full resonance, SAP of a BEC could be performed with BECs of different species, e.g., $^{85}$Rb, for which the scattering length could be tuned to almost negligible values by means of a Feshbach resonance.  

%%%%%%%%%%%%%%%%%%%%%%%%%%%%%%%%%%%%%%%%%%%%%%%%%%%%%%%%%%%%%%%%%%%%%%%%%%%%%%%%%%%%%
\section{Different non-linear interaction parameter in each well}
\label{sec:Case with multiple factors of non-linearity}
%%%%%%%%%%%%%%%%%%%%%%%%%%%%%%%%%%%%%%%%%%%%%%%%%%%%%%%%%%%%%%%%%%%%%%%%%%%%%%%%%%%%%

Let us now consider the case in which we have different non-linearities $g_i$ with $i=L,M,R$ in each well.
Note that the non-linear interaction parameter can be modified in space by the spatial variation of the scattering length using magnetic or optical Feshbach resonances \cite{Feshbach_Chin_2010, Quantum_dynamics_Clark_2015}.
In this case, the non-linear on-site energies in the $\left\{ \ket{L},\ket{M},\ket{R} \right\}$ basis are given by $\varepsilon_i=\delta_i+g_{i} |a_{i}|^2$. 
Following the same procedure described in section $\rm \ref{sec:dark_bright_basis}$,  one can derive the energies for the dark and for the dressed states in the basis $\left\{ \ket{D},\ket{+},\ket{-} \right\}$:
\begin{eqnarray}
\varepsilon_{D} &=& g_{L}\cos^4\theta+\delta_{R}\sin^2\theta+g_{R}\sin^4\theta, \label{ED3} \\
\varepsilon_{\pm} &=& \frac{1}{16}\left(\xi^{'}+8\delta_{M}\pm\sqrt{\left(\xi^{'}-8\delta_{M}\right)^2+64J^2_{\rm BM}}\right), \label{dressed_energies3}
\end{eqnarray}
where
\begin{equation}
\xi^{'}\equiv\widetilde{g}\left(1-\cos4\theta\right)+4\delta_R\left(1+\cos2\theta\right),
\label{xi'}
\end{equation}
being $\widetilde{g}\equiv g_{L}+g_{R}$.
Note that for the case $g_{L}=g_{R}=g\Rightarrow\widetilde{g}=2g\Rightarrow\xi^{'}=2\xi$ and, therefore, $\eqref{dressed_energies3}$ reduces to $\eqref{dressed_energies}$. Also for this case, one obtains $\eqref{ED}$ from $\eqref{ED3}$ after applying some trigonometric identities.

We now derive the ${\rm Ci}$ and ${\rm Cf}$ curves corresponding to the conditions $\varepsilon_{\pm}(t_{i})=\varepsilon_{D}(t_{i})$, and $\varepsilon_{\pm}(t_{f})=\varepsilon_{D}(t_{f})$:
\begin{eqnarray}
{\rm Ci}\rightarrow\;\delta_{R} &=& g_{L}+\frac{J_{\rm MR}^2(t_i)}{2(\delta_M-g_{L})},
\label{cicf1}\\
{\rm Cf}\rightarrow\;\delta_{R} &=& -g_{R}+\frac{1}{2}\left(\delta_M\pm\sqrt{J_{\rm LM}^2(t_f)+\delta_M^2}\right),
\label{cicf2}
\end{eqnarray}
with $J_{\rm MR}^2(t_i),J_{\rm LM}^2(t_f)\rightarrow0$.

%%%%%%%%%%%%%%%%%%%%%%%%%%%%%%%%%%%%%%%%%%%%%%%%%%%%%%%%%%%%%%%%%%%%%%%%%%%%%%%%%%%%%
\begin{figure*}[t!]
\centering
\includegraphics[width=1\textwidth]{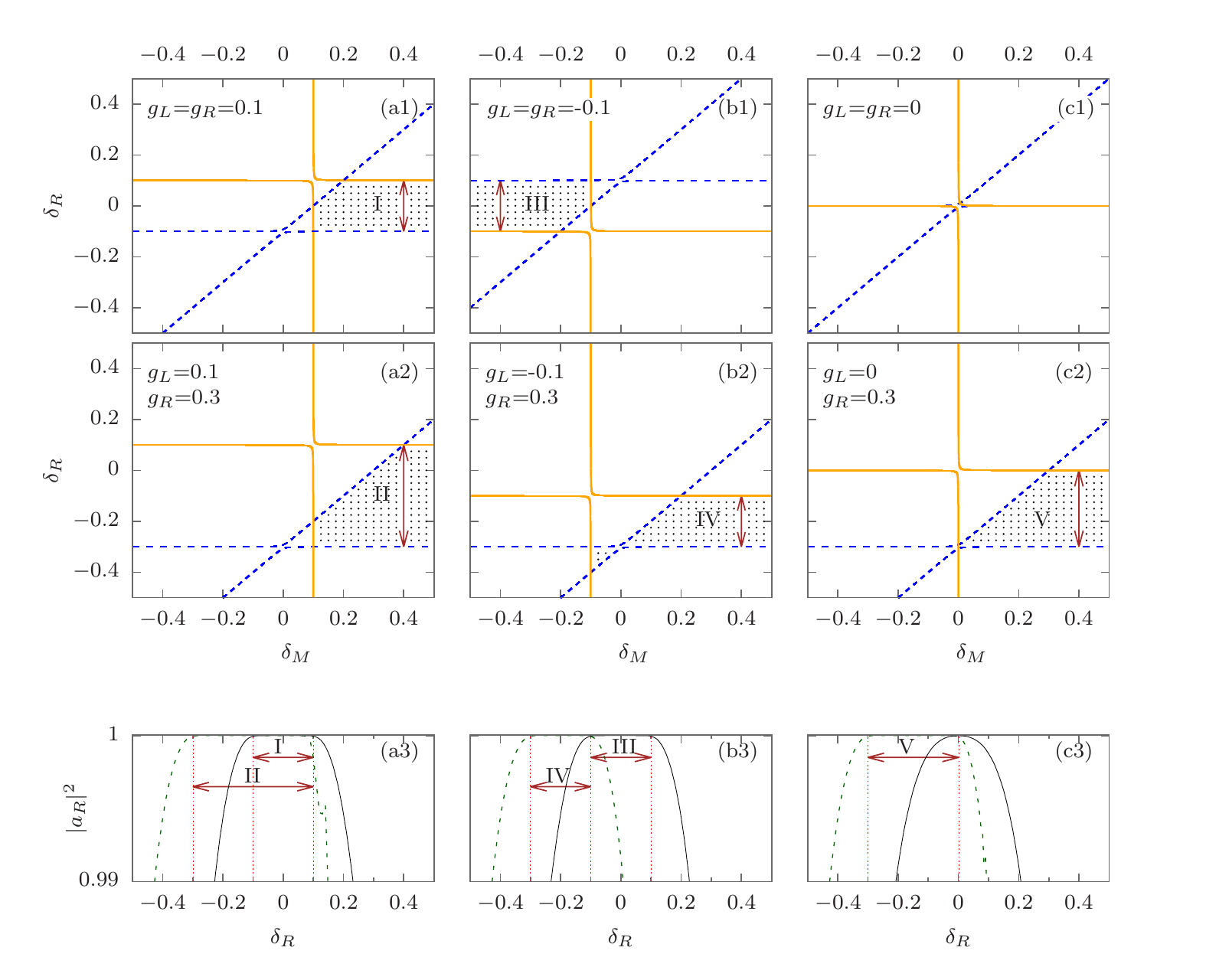}
\caption{
OZ for the SAP of a BEC (black dotted pattern) delimited by curves Ci (solid orange lines) and Cf (dashed blue lines) for equal (a1, b1, c1) and for different (a2, b2, c2) non-linear parameters in each well.
Third row: Efficiency for the SAP of a BEC from \ket{L} to \ket{R} as a function of $\delta_{R}$ for the above scenarios and a fixed value of $\delta_M$ with the red horizontal arrows indicating the extension of the plateaus. The width of these plateaus perfectly agrees with the width of the corresponding OZ in the upper figures highlighted with red vertical arrows.  
Note that in (c3) the efficiency curve for $g_{L}=g_{R}=0$ (dashed green line) does not posses a plateau. In all cases $g_{M}=(g_{L}+g_{R})/2$.  
The temporal variation of the tunnelings is the same as in Fig.~\ref{fig:3}(d).
Parameters and variables in h.o. units.
}
\label{fig:7}
\end{figure*}
%%%%%%%%%%%%%%%%%%%%%%%%%%%%%%%%%%%%%%%%%%%%%%%%%%%%%%%%%%%%%%%%%%%%%%%%%%%%%%%%%%%%%
The intersection of both regions delimited by the curves Ci and Cf, OZ$(t_i)\cap$OZ$(t_f)$, defines the OZ, which is characterized by the following inequalities.

For all cases,
\begin{subequations}
\label{TZ inequality20}
\begin{eqnarray}
\label{TZ inequality21}
\delta_{M}>g_{L},\;\; \textrm{for}\;\; g_{R}>0, \\
\label{TZ inequality22}
\delta_{M}<g_{L},\;\; \textrm{for}\;\; g_{R}<0.
\end{eqnarray}
\end{subequations}

In addition,

i) for $g_{R},g_{L}>0$ we have
\begin{subequations}
\label{TZ inequality23}
\begin{eqnarray}
\label{TZ inequality24}
-g_{R}<\delta_{R}<g_{L}, \\
\label{TZ inequality25}
\delta_{R}<\delta_{M}-g_{R},
\end{eqnarray}
\end{subequations}
and the reverse inequalities for $g_{R},g_{L}<0$,

ii) for $g_{R}<0$, $g_{L}>0$ and $g_{L}<-g_{R}$ we have two regions: 
\begin{subequations}
\label{TZ inequality26}
\begin{eqnarray}
\label{TZ inequality27}
\delta_{M}<0, \\
\label{TZ inequality28}
g_{L}<\delta_{R}<-g_{R}, \\
\label{TZ inequality29}
\delta_{R}>\delta_{M}-g_{R}
\end{eqnarray}
\end{subequations}
and
\begin{subequations}
\label{TZ inequality30}
\begin{eqnarray}
\label{TZ inequality31}
\delta_{M}>0, \\
\label{TZ inequality32}
-g_{R}<\delta_{R}<g_{L}-g_{R}, \\
\label{TZ inequality33}
\delta_{R}<\delta_{M}-g_{R},
\end{eqnarray}
\end{subequations}

iii) for $g_{R}<0$, $g_{L}>0$ and $g_{L}>-g_{R}$ we have: 
\begin{subequations}
\label{TZ inequality34}
\begin{eqnarray}
\label{TZ inequality35}
g_{L}<\delta_{R}<g_{L}-g_{R}, \\
\label{TZ inequality36}
\delta_{R}<\delta_{M}-g_{R}.
\end{eqnarray}
\end{subequations}

Finally, for the opposite cases $g_{R}>0$, $g_{L}<0$ and $g_{R}>-g_{L}$ we have the reverse inequalities of $\eqref{TZ inequality26}$-$\eqref{TZ inequality30}$, and for $g_{R}>0$, $g_{L}<0$ and $g_{R}<-g_{L}$ we have the reverse inequalities of $\eqref{TZ inequality34}$.

Following the same lines as for the case of identical non-linear parameter $g$ in each well discussed in Section~$\rm \ref{sec:optimal_conditions}$, it can be shown that the previous conditions $\eqref{TZ inequality20}$-$\eqref{TZ inequality34}$ imply $\eqref{TZ condition}$ (see Appendix B for details).

Fig.~$\ref{fig:7}$ shows numerical simulations for the SAP of a BEC with a different non-linear interaction parameter in each well. In all cases, the middle well non-linear coupling has been taken as $g_M=(g_L+g_R)/2$.
Three different cases depending on the sign of the product $g_{L}g_{R}$ have been examined.

\textsl{Case $g_{L}g_{R}>0$.}
In Figs.~$\ref{fig:7}$(a1) and (a2) the OZ (dotted pattern) for $g_{R}=g_{L}=0.1$ and for $g_{R}=3g_{L}=0.3$ have been represented, respectively. The efficiency curves for the SAP of a BEC for both cases with $\delta_{M}=0.4$ are represented in Fig.~$\ref{fig:7}$(a3). Note that for $g_{R}\neq g_{L}$ (dashed green curve) the plateau is twice larger than for the case with the same non-linear parameter in all wells (solid black curve). For the case $g_{L}<0$ and $g_{R}<0$, the OZ can be obtained performing two specular reflection, first respect to the axis $\delta_{R}$ and then respect to the axis $\delta_{M}$.

\textsl{Case $g_{L}g_{R}<0$.}
In Figs.~$\ref{fig:7}$(b1) and (b2) the OZ for $g_{R}=g_{L}=-0.1$ and for $g_{R}=-3g_{L}=0.3$ have been represented, respectively.
The SAP efficiency in both cases has been plotted in Fig.~$\ref{fig:7}$(b3) where the plateaus III and IV indicated in Figs.~$\ref{fig:7}$(b1) and (b2) are highlighted with double red arrows for $\delta_{M}=-0.4$ (solid black curve) and $\delta_{M}=0.4$ (dashed green curve), respectively.
For the case with opposite signs, $g_{L}>0$ and $g_{R}<0$, the results are a specular reflection first respect to the axis $\delta_{R}$ and then respect to the axis $\delta_{M}$.

\textsl{Case $g_{L}g_{R}=0$.}
In Figs.~$\ref{fig:7}$(c1) and (c2) the OZ (dotted pattern) for 
$g_L=g_R=0$, and $g_L=0$ with $g_R=0.3$ have been plotted, respectively. Note that the case $g_{L}=g_{R}=0$ corresponds to the Fig.~$\ref{fig:4}$(b) in which the OZ is almost negligible. The SAP efficiency has been plotted in Fig.~$\ref{fig:7}$(c3) showing the absence of a plateau for $g_L=g_R=0$ with $\delta_{M}=0$ (solid black curve) and a relatively wide plateau denoted as V for $g_{L}=0,g_{R}=0.3$ with $\delta_{M}=0.4$ (dashed green curve) as predicted in Fig.~$\ref{fig:7}$(c2).
For the case with opposite signs, the results are a specular reflection first respect to the axis $\delta_{R}$ and then respect to the axis $\delta_{M}$. 

As shown in Fig.~$\ref{fig:7}$, the ability to locally modify the non-linear interaction parameter leads, for certain parameter values, to an increase of the width of the efficiency plateaus compared to the case with the same non-linear parameter in all the wells.

%%%%%%%%%%%%%%%%%%%%%%%%%%%%%%%%%%%%%%%%%%%%%%%%%%%%%%%%%%%%%%%%%%%%%%%%%%%%%%%%%%%%%
\section{Conclusions}
\label{sec:conclusions}
%%%%%%%%%%%%%%%%%%%%%%%%%%%%%%%%%%%%%%%%%%%%%%%%%%%%%%%%%%%%%%%%%%%%%%%%%%%%%%%%%%%%%

We have obtained the optimal conditions to perform SAP of a BEC in a triple-well potential, taking into account the BEC non-linear parameter and the on-site energy in each well. 
Transforming the  three-mode Hamiltonian of the system from the bare basis to the dark/dressed basis, we have analytically derived the energies of the dark and dressed states. The resonances between the dark state and the dressed states allow to define the optimal conditions to perform SAP of a BEC. This approach is complementary to the non-linear dynamics approach discussed in Ref.~\cite{BEC_SAP_Graefe_2006}, which yields a  complete characterization of the bifurcation scenario once the adiabaticity breaks down.  

A detailed discussion of the optimal parameter region, referred as the Optimal Zone (OZ), to achieve a highly efficient SAP of the BEC has been performed. In particular, we have analytically derived the OZ boundaries and check the corresponding predictions by numerical integration of the 1D Gross--Pitaveskii equation within the three mode approximation. Worth to remark, we have demonstrated that the BEC non-linearity allows for the appearance of high-efficient transport plateaus as a function of the energy bias between the wells, which are not present in the linear case. The width of these plateaus can even be further increased by an appropriate tuning of the non-linear interaction parameter in each well.  

Finally, we would like to indicate that a detailed numerical investigation through a direct integration of the full 1D GPE still remains to be performed. However, it would be difficult to quantitatively compare the results obtained from the three-mode approximation with those obtained from the integration of the GPE, since the tunneling rates and the energy biases are independent control parameters in the former while they are dynamical variables in the latter. Note that although SAP is robust under variations of the tunneling rates it is very sensitive, particularly for a BEC, to fluctuations in the energy biases between wells. Thus, the results obtained from the three-mode approximation must be considered as a global landscape on the parameter region for which SAP for a BEC works.

%%%%%%%%%%%%%%%%%%%%%%%%%%%%%%%%%%%%%%%%%%%%%%%%%%%%%%%%%%%%%%%%%%%%%%%%%%%%%%%%%%%%%
\section*{Acknowledgments}
%%%%%%%%%%%%%%%%%%%%%%%%%%%%%%%%%%%%%%%%%%%%%%%%%%%%%%%%%%%%%%%%%%%%%%%%%%%%%%%%%%%%%

The authors acknowledge financial support through the Spanish and Catalan contracts FIS2014-57460-P and SGR2014-1639. 
JM and TB are grateful to JSPS for the Research Fellowship S-15025 and the Grant-in-Aid for Scientific Research (Grant No. 26400422), respectively. This work was also supported by the Okinawa Institute of Science and Technology Graduate University.
Finally, Albert Benseny is also acknowledged for fruitful discussions.

%%%%%%%%%%%%%%%%%%%%%%%%%%%%%%%%%%%%%%%%%%%%%%%%%%%%%%%%%%%%%%%%%%%%%%%%%%%%%%%%%%%%%
\section*{Appendix A}
%%%%%%%%%%%%%%%%%%%%%%%%%%%%%%%%%%%%%%%%%%%%%%%%%%%%%%%%%%%%%%%%%%%%%%%%%%%%%%%%%%%%%

Let us consider the case $g>0$, and show that expressions $\eqref{TZ inequality0+}$ imply $\varepsilon_{-}(\theta)<\varepsilon_{D}(\theta)<\varepsilon_{+}(\theta)$ for $0<\theta<\pi/2$.

\textit{I) Proof of $\varepsilon_{D}(\theta)<\varepsilon_{+}(\theta)$}
\hfill

To demonstrate that $\varepsilon_{D}(\theta)<\varepsilon_{+}(\theta)$, let us temporarily assume that $J_{\rm BM}=0$ in (\ref{dressed_energies}) since this coupling only increases the value of $\varepsilon_{+}(\theta)$.
By evaluating the derivative of $\xi(\theta)$, defined in $\eqref{xi}$, with respect to $\theta$, we obtain one maximum at 
$\theta_{1}=\arccos\frac{\sqrt{2g+\delta_{R}}}{2\sqrt{g}}$, yielding $\xi(\theta_{1})=\frac{\left(2g+\delta_{R}\right)^2}{2g}$.
It can be seen that
\begin{equation}
\xi(\theta_{1})=2\left(g+\delta_{R}+\delta_{R}^2/4g\right)<4\delta_{M},
\end{equation}
since $g+\delta_{R}<\delta_{M}$, and $\delta_{R}^2/4g<g^2/4g<\delta_{M}/4<\delta_{M}$, where we have used (\ref{TZ inequality3+}) and (\ref{TZ inequality2+})-(\ref{TZ inequality1+}), respectively. Note that (\ref{TZ inequality2+}) implies $\delta_{R}^{2}<g^{2}$.

As $\xi(\theta_{1})$ is a maximum, $\xi(\theta_{1})<4\delta_{M}\Rightarrow\xi(\theta)<4\delta_{M}$ within the interval.
Therefore, from (\ref{dressed_energies}), $\varepsilon_{+}(\theta)=\delta_M$ for $J_{BM}=0$ and $\varepsilon_{+}(\theta)>\delta_M$ for $J_{BM} \neq 0$.

On the other hand, $\varepsilon_{D}(\theta)$ has two maxima at $\theta_{2}=0$ and $\theta_{3}=\pi/2$, giving 
\begin{subequations}
\label{demo1}
\begin{eqnarray}
\label{demo1a}
\varepsilon_{D}(\theta_{2}) &=& g, \\
\label{demo1b}
\varepsilon_{D}(\theta_{3}) &=& g+\delta_{R}. 
\end{eqnarray}
\end{subequations}
Thus, according to (\ref{TZ inequality1+}) and (\ref{TZ inequality3+}), $\varepsilon_{D}(\theta)<\varepsilon_{+}(\theta)$.

\textit{II) Proof of $\varepsilon_{-}(\theta)<\varepsilon_{D}(\theta)$}
\hfill

$\varepsilon_{-}(\theta)$ has a maximum and $\varepsilon_{D}(\theta)$ has a minimum at {$\theta_{1}$.
Imposing $\varepsilon_{-}(\theta_{1})=\varepsilon_{D}(\theta_{1})$, it is possible to find a threshold value for the tunneling amplitude, $J_{0,\rm{min}}$, above which the crossing between these two functions can be avoided. In general, assuming $g\neq0$,
\begin{equation}
\label{J0bis}
J_{0,\rm{min}}=\cfrac{1}{2}\left|\cfrac{\delta_{R}}{g}\right|\sqrt{\cfrac{\delta_{R}^2}{2}+4g\delta_{M}-2g\delta_{R}-2g^2}.
\end{equation}
An analogous proof can be made for the case $g<0$.

%%%%%%%%%%%%%%%%%%%%%%%%%%%%%%%%%%%%%%%%%%%%%%%%%%%%%%%%%%%%%%%%%%%%%%%%%%%%%%%%%%%%%
\section*{Appendix B}
%%%%%%%%%%%%%%%%%%%%%%%%%%%%%%%%%%%%%%%%%%%%%%%%%%%%%%%%%%%%%%%%%%%%%%%%%%%%%%%%%%%%%

Let us consider the case $g_{L},g_{R}>0$ for which $\eqref{TZ inequality21}$ and $\eqref{TZ inequality23}$ apply.

\textit{I) Proof of $\varepsilon_{D}(\theta)<\varepsilon_{+}(\theta)$}
\hfill

To demonstrate that $\varepsilon_{D}(\theta)<\varepsilon_{+}(\theta)$, let us temporarily assume that $J_{\rm BM}=0$ in (\ref{dressed_energies3}) since this coupling only increases the value of $\varepsilon_{+}(\theta)$. 
By evaluating the derivative of $\xi'(\theta)$, defined in $\eqref{xi'}$, with respect to $\theta$, we obtain one maximum at 
$\theta_{4}=\arccos\sqrt{(\widetilde{g}+\delta_{R})/2\widetilde{g}}$, yielding $\xi'(\theta_{4})=2\left(\widetilde{g}+\delta_{R}\right)^2/\widetilde{g}$.
It can be seen that $\xi'(\theta_{4})<8\delta_{M}$ which is a consequence of
\begin{equation}
\label{xxx}
2\left(\widetilde{g}+2\delta_{R}+\delta_{R}^2/\widetilde{g}\right)<8\delta_{M}.
\end{equation}
To prove (\ref{xxx}) we see that $g_{R}+g_{L}+2\delta_{R}+\delta_{R}^2/\widetilde{g}<g_{L}-g_{R}+2\delta_{M}+\delta_{R}^2/\widetilde{g}<3\delta_{M}+\delta_{R}^2/\widetilde{g}-g_{R}$, where we have used (\ref{TZ inequality25}) and (\ref{TZ inequality21}) in the first and second inequality, respectively.
For $g_{R}>g_{L}$, the last two terms satisfied $\delta_{R}^2/\widetilde{g}-g_{R}<g_{R}^2/\widetilde{g}-g_{R}=-g_{L}g_{R}/\widetilde{g}<g_{L}<\delta_{M}$, and, for $g_{R}<g_{L}$, we have that $\delta_{R}^2/\widetilde{g}-g_{R}<g_{L}^2/\widetilde{g}-g_{R}<g_{L}-g_{R}<\delta_{M}-g_{R}<\delta_{M}$, where we have used (\ref{TZ inequality24}) and (\ref{TZ inequality25}).

As $\xi'(\theta_{4})$ is a maximum, $\xi'(\theta_{4})<8\delta_{M}\Rightarrow\xi'(\theta)<8\delta_{M}$.
Then, using $J_{\rm BM}=0$ in \eqref{dressed_energies3} gives $\varepsilon_{+}(\theta)=\delta_{M}$, been constant in the interval $(0, \pi/2)$. Therefore, we only need to be sure that the only extreme $\varepsilon_{D}(\theta_{5})=-\cfrac{\delta_{R}^2-4g_{L}(g_{R}+\delta_{R})}{4\widetilde{g}}$, where $\theta_{5}=\arccos{\sqrt{(2 g_{R}+\delta_{R})/2\widetilde{g}}}$, takes a lower value than $\delta_{M}$.
Indeed, it can be seen that
\begin{equation}
-\cfrac{\delta_{R}^2}{4\widetilde{g}}+\cfrac{g_{L}}{\widetilde{g}}\left(g_{R}+\delta_{R}\right)<\delta_{M},
\end{equation}
using $\eqref{TZ inequality25}$.}

\textit{II) Proof of $\varepsilon_{-}(\theta)<\varepsilon_{D}(\theta)$}
\hfill

To find a threshold value for the tunneling amplitude, $J_{0,\rm{min}}$, above which the crossing between $\varepsilon_{D}(\theta)$ and $\varepsilon_{-}(\theta)$ can be avoided, it is sufficient to impose $\varepsilon_{D}(\theta_{5})=\varepsilon_{-}(\theta_{4})$, where $\theta_{5}$ (introduced in proof~\textit{I} above) and $\theta_{4}$ are the extreme values of $\varepsilon_{D}(\theta)$ and $\varepsilon_{-}(\theta)$, respectively, in the interval $(0,\pi/2)$. From this condition, we obtain
\begin{widetext}
\begin{equation}
J_{0,\rm{min}}=\cfrac{1}{2\left|\widetilde{g}\right|}\sqrt{\left(g_{L}^2+g_{R}^2+2g_{R}\delta_{R}+2\delta_{R}^2-2g_{L}\left(g_{R}+\delta_{R}\right)\right)(4g_{R}\delta_{M}+\delta_{R}^2-4g_{L}(g_{R}-\delta_{M}+\delta_{R}))},
\end{equation}
\end{widetext}
which becomes \eqref{J0bis} for $g_{L}=g_{R}=g$.

An analogous proof can be made for the rest of the cases corresponding to different combinations of signs of $g_{L}$ and $g_{R}$.

%%%%%%%%%%%%%%%%%%%%%%%%%%%%%%%%%%%%%%%%%%%%%%%%%%%%%%%%%%%%%%%%%%%%%%%%%%%%%%%%%%%%%

\end{document}